\documentclass[preprint,12pt]{elsarticle}

\usepackage{amssymb}
\usepackage{amsthm}%
\usepackage[fleqn]{amsmath}
\usepackage{caption}
\usepackage{subfig}
\usepackage[hyphens]{url}
\usepackage{hyperref}

\makeatletter
\def\ps@pprintTitle{%
  \let\@oddhead\@empty
  \let\@evenhead\@empty
  \def\@oddfoot{\reset@font\hfil\thepage\hfil}
  \let\@evenfoot\@oddfoot
}
\makeatother

\journal{Nuclear Instruments \& Methods in Physics Research A. }

\begin{document}

\begin{frontmatter}



\title{Coincidence-based reconstruction for reactor antineutrino detection in gadolinium-doped Cherenkov detectors}


\author[inst1]{L. Kneale\corref{cor1}}
\cortext[cor1]{Corresponding author: e.kneale@sheffield.ac.uk}
\affiliation[inst1]{organization={Department of Physics \& Astronomy, University of Sheffield},
            addressline={Hicks Building, Broomhall}, 
            city={Sheffield},
            postcode={S3 7RH}, 
            country={United Kingdom}}

\author[inst2]{M. Smy}
\author[inst1]{M. Malek}

\affiliation[inst2]{organization={Department of Physics \& Astronomy, University of California Irvine},
            addressline={Frederick Reines Hall}, 
            city={Irvine},
            postcode={92697-4575}, 
            state={California},
            country={USA}}

\begin{abstract}
A reconstruction algorithm has been developed to capitalize on advances in Cherenkov technology for reactor antineutrino detection. 

Large gadolinium-doped water (Gd-H$_2$O) Cherenkov detectors are a developing technology which use Gd loading to increase the visibility of the neutrons produced in inverse beta decay (IBD) interactions, which produce positron-neutron pairs coincident in time and space. In this paper, we describe the reconstruction which uses the combined light from both events in an IBD pair to accurately reconstruct the interaction vertex. 

Using simulation, the algorithm has been applied to the reconstruction of reactor antineutrinos in Gd-H$_2$O and in Gd-doped water-based liquid scintillator (Gd-WbLS), an advanced detector medium which is also currently in development.

Compared to a single-event reconstruction, the combined reconstruction improves vertex resolution for reactor IBD positrons by up to a factor of 4.5 at the lowest energies. IBD-neutron vertex resolution was found to improve by more than 30\% in most instances. 

Powerful background rejection with the coincidence reconstruction can be achieved by requiring a minimum quality of fit. This was found to reject up to 94\% of accidental coincidences of uncorrelated background events, while retaining at least 97.5\% of the IBD signal pairs.

\end{abstract}




\end{frontmatter}


\section{Introduction}

As Reines and Cowan showed\cite{Cowan1956}, the antineutrino emission from a reactor can be detected via the inverse $\rm \beta$ decay (IBD) weak interaction of antineutrinos with free protons in water or a hydrocarbon liquid: 
$$\rm \overline{\nu}_e + p \longrightarrow e^+ + n.$$
This is the principal interaction of the antineutrino at the low energies of reactor antineutrinos.

A nascent water Cherenkov technology - gadolinium (Gd) doping - opens up the possibility of detecting reactor antineutrinos in a water or water-based Cherenkov detector. Large Gd-doped water (Gd-H$_2$O) Cherenkov detectors use Gd loading to tag the neutrons produced in the IBD interaction. The Gd-H$_2$O technology was first demonstrated in \cite{Marti2020} and other detectors have more recently followed suit~\cite{Abe2021,back2017}.

In pure water, the IBD neutron captures on hydrogen and the low light yield makes this difficult to observe. In Gd-H$_2$O, the neutron captures preferentially on gadolinium at concentrations greater than 0.01\%, and this increases the light yield from the capture of the IBD neutron by a factor of 3 to 4. In addition, the coincidence of the neutron capture with the positron signal is closer in time than in pure water, which enhances background rejection.

Liquid scintillator detectors are a proven technology for reactor antineutrino detection~\cite{Gando2011} and Gd-doped scintillator detectors benefit from the increased light yield and the coincidence of the neutron capture close in distance and time to the positron vertex~\cite{Abe2012,An2013,Ahn2012}.

Water-based liquid scintillator (WbLS)~\cite{Yeh2011} is an emerging medium, which has the potential to combine the higher light yield and lower-energy sensitivity of scintillation detectors with the directional information and large scale of water Cherenkov detectors with benefits for reactor antineutrino detection~\cite{Askins2020,Zsoldos2022}.

The accuracy of vertex reconstruction is important for reducing systematic error on the definition of the fiducial volume, for background rejection and for reconstruction of the antineutrino energy. Improvements at lower energies in particular can improve overall sensitivity to reactor antineutrinos and help to lower the energy threshold of a detector.

The displacement of the neutron capture from the primary IBD interaction vertex is small compared to the vertex resolution in Gd-doped detectors - with a mean distance of $\sim6$~cm, over 90\% of neutrons capture within 10~cm and the remaining capture within 35~cm. The neutron-capture point itself does not emit light but the Gd emits a number of gammas and the Cherenkov light resulting from Compton scatters of these gammas can be used to reconstruct an ‘effective’ neutron-capture vertex.

A novel reconstruction algorithm, which capitalizes on the spatial coincidence of the positron and neutron-capture signal pair in the emerging Gd-doping technology, has been developed and applied to interactions of antineutrinos in the reactor spectral range, using Monte Carlo simulations.

This paper describes the \textit{coincidence reconstruction} that has been implemented specifically to reconstruct the position of events in a Gd-doped detector medium, by reconstructing pairs of events together. Section~\ref{sec:reacantinu} describes the fundamentals of reactor antineutrino detection in Gd-doped media and Section \ref{sec:sims} describes the Monte Carlo simulations used in the remainder of the paper. Section \ref{reco:bonsai} describes the established maximum likelihood fitter for single-event reconstruction which forms the basis of the coincidence reconstruction. Section \ref{reco:pairbonsai} details the extension of this fitter to a coincidence reconstruction and its implementation for reactor antineutrinos. Improvements to vertex resolution and event selection/rejection in the reactor antineutrino energy range are presented and discussed in Sections~\ref{core:vertex-res} and~\ref{core:tagging} for two Gd-doped Cherenkov detection media in two different-sized detectors. Conclusions are drawn in Section~\ref{sec:conclusions}. Some of the material included in this paper has been taken from~\cite{kneale2021}.

\section{Reactor antineutrinos in gadolinium-doped Cherenkov detectors}
\label{sec:reacantinu}
 
In a Cherenkov detector, positrons from the IBD interaction with a total energy above the Cherenkov threshold of $\sim$0.8~MeV emit a prompt signal. The detectable spectrum of the IBD positrons resulting from reactor antineutrino interactions is in the range $\sim$0.8~MeV to $\sim$8~MeV total energy, with a peak at $\sim$2.4~MeV, given a peak reactor antineutrino energy for IBD of $\sim$3.7~MeV\cite{Vogel2015}. The neutrons from the IBD thermalize and are captured on nuclei in the medium, emitting a delayed signal. In pure water, the IBD neutrons capture on hydrogen, resulting in the delayed emission of a single 2.2~MeV gamma as the resulting deuteron decays to ground state. This occurs within a mean time of $\sim$200~$\rm \mu s$ of the prompt signal.

The principle of using Gd-H$_2$O for low-energy reactor antineutrino detection was first suggested by~\cite{Bernstein2001}. The neutron captures preferentially onto the Gd due to the very high neutron-capture cross section of Gd ($\sim$49,000~b) compared to hydrogen ($\sim$0.3~b). At a concentration of 0.1\% Gd ions, which can be achieved with the addition of 0.2\% gadolinium sulfate, 90\% of the neutrons may capture onto Gd~\cite{Beacom2003}. The remaining neutrons capture onto the hydrogen or sulfate. The subsequent decay of the Gd to ground state releases a cascade of gammas totaling $\sim$8~MeV in energy. These further interact in the water to produce Cherenkov light and the neutron-capture signal can be detected with a peak visible energy of around 4.5~MeV in a Gd-doped water Cherenkov detector, which is generally higher in energy than the positron signal. In Gd-H$_2$O, the delayed neutron-capture signal occurs within a shorter mean time of $\rm \sim$30~$\rm \mu s$.

This combination of the prompt positron and higher-energy delayed neutron-capture signal within a short space and time results in a more easily detectable correlated signal in Gd-H$_2$O compared to in pure water. This results in lower-energy sensitivity and makes the prospect of reactor antineutrino detection in a water Cherenkov detector feasible. 

The addition of a scintillating component to Gd-H$_2$O, in the form of a water-based liquid scintillator, could combine the benefits of Gd-H$_2$O, including the coincident signal pair from the Gd doping and the directional information and particle identification capabilities of Cherenkov light~\cite{kasuga1996}, with the higher light yield of scintillator detectors, for detection of reactor antineutrinos at the lowest end of the energy range.

WbLS cocktails have been developed using PPO (2,5-diphenyl-oxazole) as the wavelength-shifting scintillator component in a linear alkylbenzene (LAB) solvent~\cite{Yeh2011}. The oily scintillator component is then combined with pure water using a surfactant which creates micelles which have both hydrophilic (polar) and hydrophobic (non-polar) surfaces. Gd-doped WbLS (Gd-WbLS) is under development.

IBD positrons are emitted largely isotropically and the prompt signal comes from the single positron. Neutrons from the IBD interaction are generally emitted in the forward direction compared to that of the incoming antineutrino, although this directional information is lost within a couple of scatters as the neutron thermalizes in the medium. The light from the neutron capture on gadolinium is composed of multiple gammas in multiple directions, which results in a more isotropic light distribution compared to that of the single positron in Gd-H$_2$O. In Gd-WbLS, there is an additional contribution of isotropic scintillation light in both the prompt positron and delayed neutron signal.

\section{Detector Simulations}
\label{sec:sims}

In this paper, the coincidence reconstruction is applied to simulated interactions in two different detector sizes, each with a Gd-H$_2$O and a Gd-WbLS fill. More precisely, the two fill media are:
\begin{itemize}
    \item Gd-H$_2$O with 0.2\% Gd$_2$(SO$_4$)$_3$ doping (for 0.1\% Gd concentration) and
    \item Gd-WbLS with 0.2\% Gd$_2$(SO$_4$)$_3$ doping and $\sim$100 photons per MeV WbLS (approximately 1\% of the light yield of pure LAB-based scintillator with 2g/L of the fluor, PPO, typically used in large neutrino experiments such as Daya Bay and SNO+~\cite{Beriguete2014,Andringa2016}).
\end{itemize}

\begin{figure}
    \centering
    \includegraphics[width=0.5\linewidth]{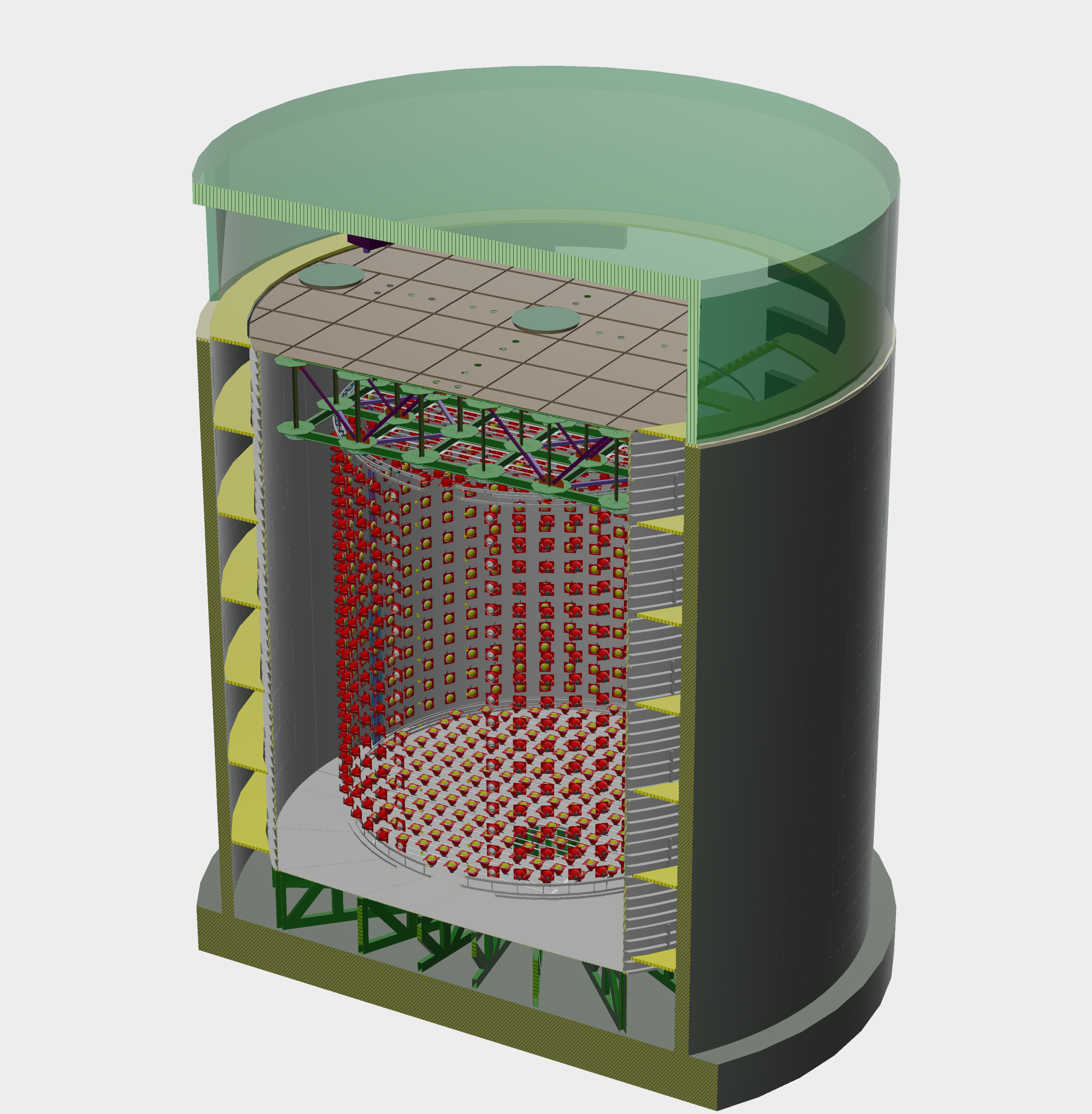}
    \caption{Schematic of the detector geometry by Jan Boissevain (University of Pennsylvania), showing the tank supported on a steel truss structure, and the inner support structure instrumented with photomultiplier tubes to create an instrumented inner volume.}
    \label{fig:detector_schematic}
\end{figure}

The two detectors are upright cylinders, with an inner PMT support structure which creates an instrumented inner detector volume within the tank. A schematic of the detector geometry is given in Figure~\ref{fig:detector_schematic} and the detector parameters are summarized in Table~\ref{tab:comp-det}. The inner volume is instrumented with Hamamatsu R7081 10 inch PMTs.

\begin{table}[htb]
    \centering
    \captionsetup{font=small}
    \caption{Summary of detector geometries used in this paper.}
 \begin{tabular}{c|c|c}
    { Tank diameter}  & { Inner }         & { Inner PMT }   \\
    { and height [m]} & { volume radius [m]} &                        { coverage [\%]} \\ \hline
    16                   & 5.7                                          & 15 \\ 
    22                   & 9.0                                            & 15 \\ 
    \end{tabular}
    \label{tab:comp-det}
\end{table}

Full Monte Carlo (MC) detector simulations were carried out with an adaptation of RAT-PAC (Reactor Analysis Tool - Plus Additional Codes) \cite{Seibert2014}, which is based on the physics simulation framework GEANT4 \cite{ALLISON2016,AGOSTINELLI2003}, the CLHEP physics library \cite{LONNBLAD1994}, the GLG4sim (Generic Liquid-scintillator Anti-Neutrino Detector or \textit{GenericLAND}) Geant4 simulation for neutrino physics \cite{glg4sim} and the data analysis framework ROOT \cite{BRUN1997}.

The MC model for WbLS is detailed in~\cite{Land2021}. The time profile of scintillation light is based on measurements of 1\% WbLS~\cite{Onken2020}, with a fast rise time of 0.25~ns and a prompt decay time of 2.01~ns. This timing profile leads to a large overlap of Cherenkov and scintillation light, which is beneficial for vertex resolution since it adds light while smearing the timing only minimally. The light yield consistent with 1\% of the light yield from pure LS (100 photons per MeV) and scattering were taken from measurements of Gd-WbLS~\cite{Gabriel2022}. The time profile measured for the same Gd-WbLS cocktail was found to be consistent with that of the unloaded WbLS time profile, within error.

\section{Low-energy single-event reconstruction}
\label{reco:bonsai}
{
The single-event reconstruction - BONSAI (Branch Optimization Navigating Several Annealing Iterations) - was originally written to reconstruct low-energy events from Cherenkov light in water Cherenkov detectors and has been used for many years in Super-Kamiokande for reconstruction of events up to 100~Mev~\cite{Smy2007}.

BONSAI is a point fitter, that is it assumes that all light originates from a single point in space. BONSAI reconstructs the origin point of the positron near the IBD interaction vertex from the Cherenkov light emitted as it travels through the detector medium at relativistic speeds. The gammas emitted by the neutron capture produce relativistic electrons and positrons through Compton scattering or pair production. These emit Cherenkov cones as they travel through the detector and BONSAI reconstructs the neutron vertex as the point source which best fits all of the Cherenkov light from the multiple gammas resulting from the neutron capture.

It is a maximum likelihood fitter to the PMT hit timing. The likelihood is based on the hit time residuals of the Cherenkov signal in Gd-H$_2$O (or Cherenkov + scintillation signal in Gd-WbLS) and dark noise background. It is calculated for a selection of test vertices and is given by:
\begin{equation}
    \rm ln\mathcal{L}(\boldsymbol{x},t_0) = ln(\prod_{i=1}^{N} \: P(\Delta t_i (\boldsymbol{x}))).
    \label{equ:likelihood}
\end{equation}
The hit time residual $\rm \Delta t_i (\boldsymbol{x})$ is:
\begin{equation}
        \rm \Delta t_i (\boldsymbol{x}) = t_i - tof_i (\boldsymbol{x})-t_0
\end{equation}
where $\rm \boldsymbol{x}$ is the test vertex, $\rm t_i$ is the hit time at the i$^{th}$ PMT, $\rm t_0$ is the emission time and $\rm tof_i = |\boldsymbol{x_i}-\boldsymbol{x}|/c_{medium}$ is the time of flight from the reconstructed vertex to the PMT vertex for hit i and $c_{medium}$ is the group velocity of light, averaged over the Cherenkov and scintillation light spectrum, specific to the detector medium. This was determined from simulation for both media for the purposes of this paper.

The dark noise component is calculated by taking the rate of hits outside the signal window and scaling it to the size of the signal window. The signal window is $\rm t_i - tof_i(\boldsymbol{x})$ for a given test vertex $\rm \boldsymbol{x}$.

$P(\Delta t_i)$ is a probability density function (PDF) which is defined using hit time residuals from true vertices in calibration data or Monte Carlo (MC) simulation. The timing residual PDFs fold in the effects of PMT timing features, photocoverage and scattering and reflection in the detector medium but do not depend directly on the light's angle of incidence, distance traveled or on the location of the PMTs. In Figure \ref{fig:time_residuals}, which shows the timing residuals used for the detector configurations under analysis, the shape is dominated by the PMT timing features discussed for a similar photomultipler tube in \cite{Brack2013} with the prompt peak at zero, the double-pulsing peaking at $\sim$50~ns and the late-pulsing peak at $\sim$70~ns. The key parameter affecting the reconstruction is the transit-time spread (TTS) of the PMT, which is the $\sigma$ value of the prompt peak. For the Hamamatsu R7081 PMT, this is around 2~ns. For PMTs with a larger transit-time spread, the prompt peak is wider and this can make the fit less accurate. Scattering in increasing tank sizes results in increasing tails out to longer times. The addition of liquid scintillator in Gd-WbLS results in a wider prompt peak due to absorption and re-emission by the scintillator and consequently less well-defined prompt and double-pulsing peaks. Other features in the PDFs such as the timing and quantity of after pulsing also impact the reconstruction. The PDFs of time residuals are derived from simulation for this paper.

\begin{figure}[htb]
\centering
    \includegraphics[width=\textwidth]{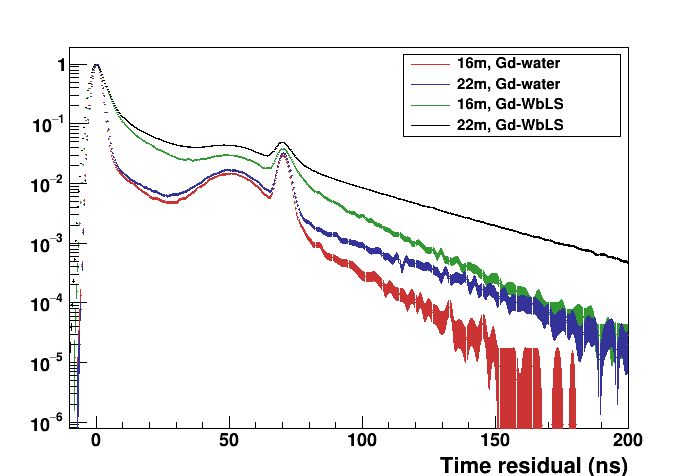}
  \captionsetup{width = \textwidth,font=small}
    \caption{PDFs of true hit-time residuals calculated from the MC vertex using simulation for the two detector sizes, each with a Gd-H$_2$O and Gd-WbLS fills. These show the effects of PMT timing, increasing scattering with detector size and wider peaks with the addition of scintillator.}
  \label{fig:time_residuals}
\end{figure}
For a given triggered event, hits are first passed through a hit selection criterion, which creates a list of hits that can be used to generate a sample of vertices which form the starting point for the likelihood maximization. This is done by removing isolated hits and then requiring that for any one pair of PMT hits separated by time $\Delta \mathit{t}$, the distance that could be traveled by direct light in the time between the hits ($\frac{\mathit{c}}{\mathit{n}} \Delta \mathit{t}$) is less than the distance between the two hit PMTs. This ensures that in principle the light is unscattered and could have come from the same interaction.

A minimum of four hits is required to reconstruct a vertex in 3D space. Sets of four hits are selected from the list of direct hits and used to define a test vertex by solving all four equations for the hit arrival times simultaneously, exactly and analytically for x, y, z and t$_0$. In this way, each quadruple of hits defines a point in the detector and a list of potential initial test vertices for the maximum likelihood vertex search is generated. Having more than one starting point helps to avoid mis-reconstruction due to local maxima.

Since the number of quadruples is proportional to the number of hits $\mathit{N}^4$, some limits are applied to increase speed. The number of quadruples is reduced by giving preference to four-hit combinations with less spread in absolute time. This is done by selecting a time window containing a predetermined number of combinations and maximizing over all combinations in the window. Additional quadruples are formed by combining each hit in the time window with the three hits that immediately follow it. When test vertices for all combinations in that window have been evaluated, the number of starting points is further reduced by averaging over nearby points in steps of 60~cm and 150~cm.

From the final, reduced list of starting points, likelihood maximization, with free parameters for emission time and dark noise rate, is carried out for successive iterations of searches of test vertices. 

A \textit{simulated annealing} algorithm is a stochastic technique used to find the global maximum\cite{Wiki2023}. BONSAI achieves a similar effect by allowing a range of likelihoods at each stage of the maximization. At each iteration, the process selects the test vertices with the best log likelihoods in a range to take forward to the next iteration. The range is a fraction of the total range of log likelihoods for all test vertices in that iteration and the fraction (and thus range) is reduced in each step. By accepting solutions which may be worse than the current solution along with better solutions, simulated annealing is designed to avoid local minima. The probability of accepting a worse solution decreases with each step and the result is a gradual convergence on the global local maximum.

At each stage of the process, each selected test vertex becomes a center point for a dodecahedron and the vertices of each dodecahedron become new additional test vertices. The radius of the dodecahedrons is reduced with each iteration. The dodecahedron grid shape was selected to give optimal coverage of the space, given that the distance to each vertex from the center is similar to the distance between vertices, while limiting the time taken for each fit.

In the final iteration, the vertex corresponding to the highest log likelihood is chosen as the reconstructed vertex. 

\section{Coincidence Reconstruction}
\label{reco:pairbonsai}

The coincidence reconstruction for event pairs in BONSAI uses the same building blocks as the standard BONSAI single-event fit. Light from both the positron and neutron events is used and a single, combined, reconstructed vertex is output for any event pair passed to the reconstruction. 

The hits from each event go through the hit selection process independently. Once the selected hits have been through the four-hit selection, a list of test vertices is output for each event. At this point, the lists of test vertices from the two events are combined into a single, larger list of test vertices which are used as starting points for the vertex search.

The hit information for each event is retained and the likelihood for each event is calculated simultaneously for each test vertex. The coincidence reconstruction is achieved via the maximization of the sum of log likelihoods for the prompt (positron or positron-like) and delayed (neutron or neutron-like) events with free parameters for prompt and delayed emission times and the dark noise rate:
\begin{equation}
    \rm ln\mathcal{L}(\boldsymbol{x},t_{0,p},t_{0,d}) = ln\mathcal{L}_p(\boldsymbol{x},t_{0,p}) + ln\mathcal{L}_d(\boldsymbol{x},t_{0,d})
\end{equation}
where the log likelihoods for the prompt and delayed events are as given in Equation \ref{equ:likelihood}. Emission times t$_{0,p}$ and t$_{0,d}$ are the fitted prompt and delayed emission times respectively.

Combining the starting solutions for each individual event into a larger list of starting solutions improves rejection of local maxima in the likelihood maximization. The addition of data points (PMT hits) is particularly helpful where light yields from one or both individual events are low. This is often the case for the positron event in the reactor antineutrino range, particularly in Gd-H$_2$O. For example, the BONSAI single-event reconstruction tends to be unstable if there are fewer than 10 inner-PMT hits, which equates to between 1 and 1.5~MeV in the Gd-H$_2$O detectors. In the 16~m (22~m) Gd-H$_2$O detector simulations used in this paper, the light yield (without dark noise) from 23\% (21\%) of the positrons produced fewer than 10 PMT hits and the average light yield was 16.72 (17.14) hits.

\subsection{CoRe Implementation of the Coincidence Reconstruction for Reactor Antineutrinos}
\label{core:implementation}

The distance of the neutron-capture vertex from the IBD interaction and positron vertex is within the expected vertex resolution and the additional light from the neutron event can therefore be used to improve the reconstruction of the positron event.

The CoRe implementation was first developed to reconstruct pairs of events using Cherenkov light in Gd-H$_2$O and optimized for the best possible vertex resolution for IBD positron-neutron pairs in this medium.
 
In BONSAI, preference is given to vertices which, when combined with the hit pattern, reflect a Cherenkov light distribution. This is achieved by correcting the log likelihood as follows:
\begin{equation}
    ln\mathcal{L}'(\textbf{x},\lambda) = ln\mathcal{L}(\textbf{x},t_0) - \frac{1}{2}\left(\frac{\theta_c-\theta_{fit}}{\sigma_\theta}\right)^2
\end{equation}
where $\theta_{c}$ (44.75\textdegree, close to the maximum Cherenkov cone opening angle for positrons in water) is the constraining angle and $\theta_{fit}$ is the opening angle calculated from the directions of the hit PMTs as seen from the vertex assuming a single Cherenkov cone. The value of $\sigma_\theta$ depends on whether the cone opening angle is less than the constraining angle (a good fit to Cherenkov light) or greater than the constraining angle (a poor fit to Cherenkov light), with values of 19.12\textdegree and 8\textdegree respectively. These were optimized in the implementation of BONSAI in Super-Kamiokande.

The prompt light in an IBD interaction originates from Cherenkov cones along the positron's track. For electrons and positrons, significant numbers of Cherenkov photons are detected only when the particles are highly relativistic and the Cherenkov cone opening angle is therefore maximal. In water, the maximum Cherenkov cone opening angle for positrons is 41.2\textdegree, although the light tends towards a diffuse circle rather than a cone due to scattering of the positron as it travels. Since the Cherenkov light from neutron-capture events results from multiple gammas, it is generally more isotropic than the Cherenkov light from the positron. Figure~\ref{fig:theta_pos_neut}a shows the anisotropy of hits from IBD positrons and neutron-capture events in the 16~m detector. The first-order Legendre polynomial coefficient

\begin{equation}
    \beta_1 = \frac{2}{N(N-1)}\sum_{i=1}^{N-1} \sum_{j=i+1}^N \cos\theta_{ij},~~~~~~~~~~~~~~~~i\neq j
\end{equation}
where \textit{N} is the number of PMT hits, is used as a measure of anisotropy, as described in~\cite{kneale2021}. 

To account for the more isotropic light from the neutrons, the angular constraint was relaxed for the CoRe implementation. The vertex resolution across the whole reactor antineutrino spectrum was calculated for a range of constraining angles and the angle which gave the best vertex resolution for the IBD pair was used. The optimal constraining angle was found to increase with the size of the detector. A 90\textdegree constraining angle gave the optimal vertex resolution in Gd-H$_2$O in the 22~m detector. The optimal vertex resolution in the 16~m detector was found to be given by a constraining angle of 80\textdegree.

\begin{figure}[htb]
    \centering
    \captionsetup[subfigure]{width=0.5\textwidth}
    \subfloat[Gd-H$_2$O 16~m detector.]{%
    \includegraphics[width=0.5\textwidth]{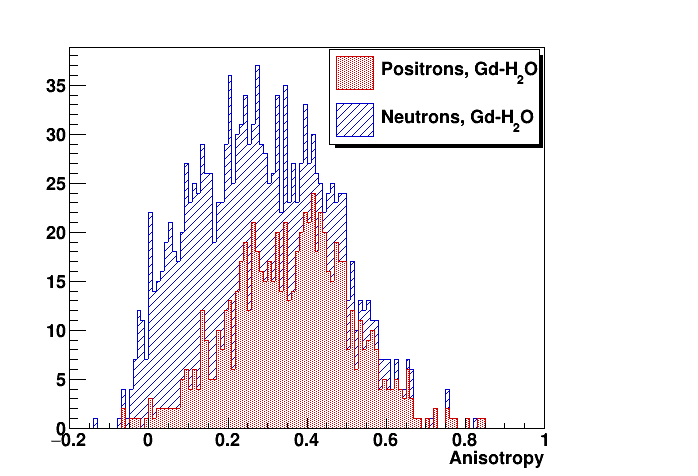}%
    }
    \subfloat[Gd-WbLS 16~m detector.]{%
    \includegraphics[width=0.5\textwidth]{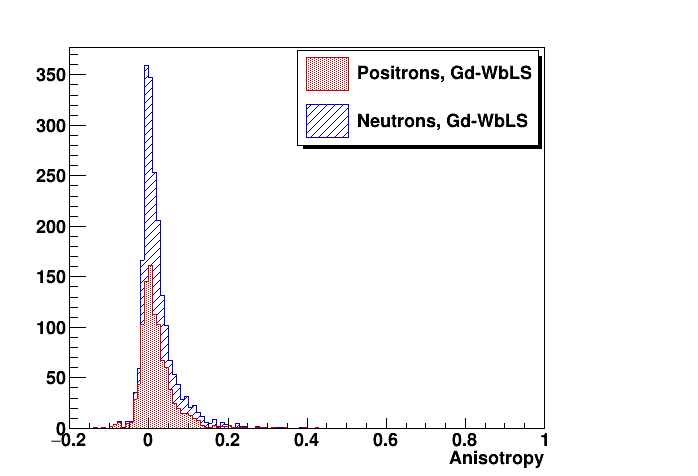}%
    }
    \caption{First-order Legendre polynomial coefficient as a measure of anisotropy of simulated hits for IBD positrons (red, shaded) and neutrons (blue, lined) in the 16~m detector. The light resulting from the neutron capture tends to be more isotropic than the light from the single positron in Gd-H$_2$O (left). In Gd-WbLS (right), the light from both the positron and neutron events is similarly isotropic.}
    \label{fig:theta_pos_neut}
\end{figure}

In order to extend the CoRe implementation to reconstruct pairs of events in Gd-WbLS, the constraint on the angle was turned off completely to allow for the highly isotropic scintillation light (Figure~\ref{fig:theta_pos_neut}b), which does not arrive with sufficient separation from the Cherenkov light to require a separate treatment. If, however, it were possible to separate the Cherenkov and scintillation light, \textit{e.g.}, by using a slower scintillator~\cite{Biller2020} or a lower concentration of PPO~\cite{Onken2020} in combination with novel light collection such as fast photosensors~\cite{Adams2016} or wavelength-based photon sorting\cite{Kaptanoglu2020}, then separate treatment of the Cherenkov and scintillation light may improve results for Gd-WbLS fills.

The results in this paper use a reconstruction threshold requiring a minimum total light yield from an event of 5 hits on the PMTs instrumenting the inner volume. Ten hits is considered to be the minimum light yield required for a reliable reconstruction in BONSAI to help reject events which reconstruct poorly. Since CoRe was expected to improve results at very low light levels, the threshold in BONSAI and CoRe was set to 5 hits for this study.

CoRe iterates over all triggers and attempts to reconstruct all pairs occurring within a specified time of each other - 200~$\mu$s in Gd-H$_2$O and 300~$\mu$s in Gd-WbLS. These loose limits on the time difference were designed to ensure that all true pairs were reconstructed while at the same time reducing computation time. The longer time limit in Gd-WbLS takes into account the wider timing distribution of the prompt and delayed triggers compared to Gd-H$_2$O. For successfully reconstructed pairs, the data output include the time between events, as well as the total charge and total number of PMT hits for each event. Additional information for each event includes a measure of the fit quality - called the \textit{timing goodness}.

\subsection{Timing Goodness - Fit Quality}
\label{core:goodness}

Where events are poorly reconstructed, the coincidence of the hit time residuals as calculated from the reconstructed vertex is also poor. BONSAI outputs the coincidence of the time residuals as a measure of the vertex fit quality - the timing goodness. The time residuals are calculated using the reconstructed time of emission, which is extracted from a fit to the peak of the time-of-flight-subtracted PMT hit times at the reconstructed vertex $\mathit{\textbf{x}}$.

More specifically, the timing goodness is given by a Gaussian distribution for the Cherenkov timing resolution, weighted by a second, wider Gaussian:
\begin{equation}
    \rm g(\boldsymbol{x}) = \frac{\sum_{hits} w_i e^{-0.5\big(\frac{\Delta t_i(\textbf{x})}{\sigma}\big)^2} }{\sum_{hits} w_i}
    \label{equ:goodness}
\end{equation}
Here, $\rm \sigma$ is the timing resolution expected for Cherenkov events and $\mathit{w_i}$ are weights based on the hit time residuals using a wider effective time resolution. The hit weights are given by a Gaussian of width $\rm \omega$:
\begin{equation}
    \rm w_i = e^{-0.5\big(\frac{ \Delta t_i (\textbf{x})}{\omega}\big)^2}
\end{equation}

The results presented in this paper for both media have timing goodness values calculated using Gaussian distributions with widths of $\sigma=4 $~ns and $\omega = 50 $~ns. An ideal reconstruction would result in timing goodness $\mathit{g}(\boldsymbol{x})=1$. 

This measure of fit quality is less well-adapted to Gd-WbLS because of the wider prompt peak in the pdf of the time residuals and the convolution of the Cherenkov and scintillation light but, for the purposes of this paper, it was found to be sufficient as a relative, rather than absolute, measure. It should be adapted to provide an accurate measure of the fit quality for reconstruction in Gd-WbLS in the future. 

}

\section{Improved Vertex Resolution with CoRe}
\label{core:vertex-res}

\begin{figure}[htb]
    \centering
    \captionsetup[subfigure]{width=0.5\textwidth}
    \subfloat[Positrons in the Gd-H$_2$O 16~m detector.]{%
    \includegraphics[width=0.5\textwidth]{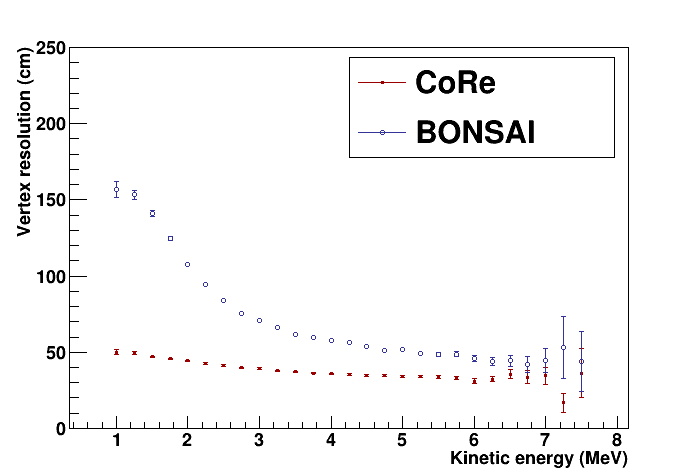}%
    }
    \subfloat[Positrons in the Gd-H$_2$O 22~m detector.]{%
    \includegraphics[width=0.5\textwidth]{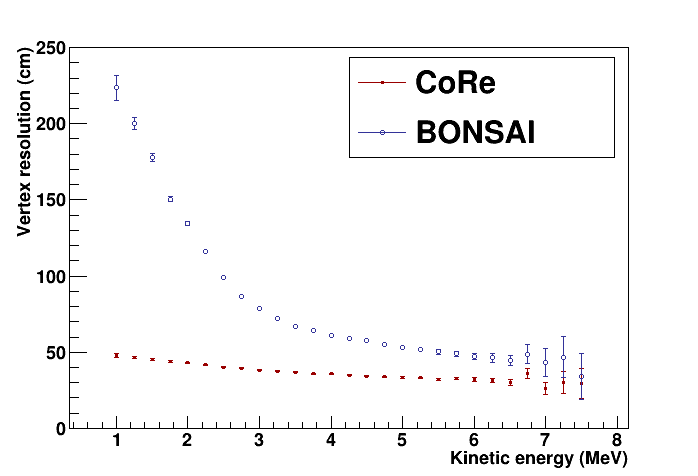}%
    }
    \captionsetup[subfigure]{width=0.5\textwidth}
    \subfloat[Neutrons in the Gd-H$_2$O 16~m detector.]{%
    \includegraphics[width=0.5\textwidth]{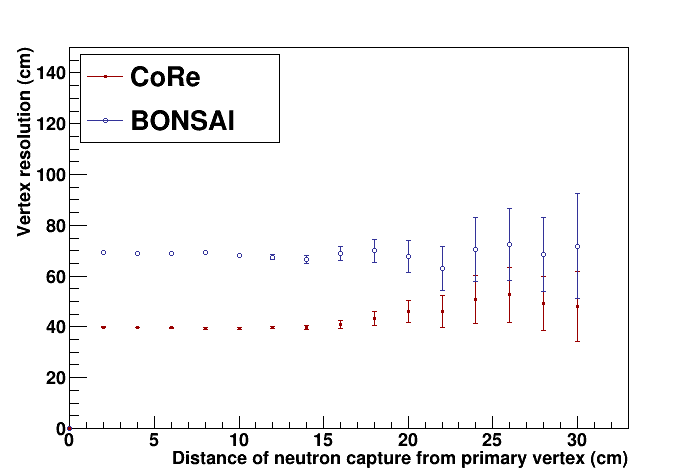}%
    }
    \subfloat[Neutrons in the Gd-H$_2$O 22~m detector.]{%
    \includegraphics[width=0.5\textwidth]{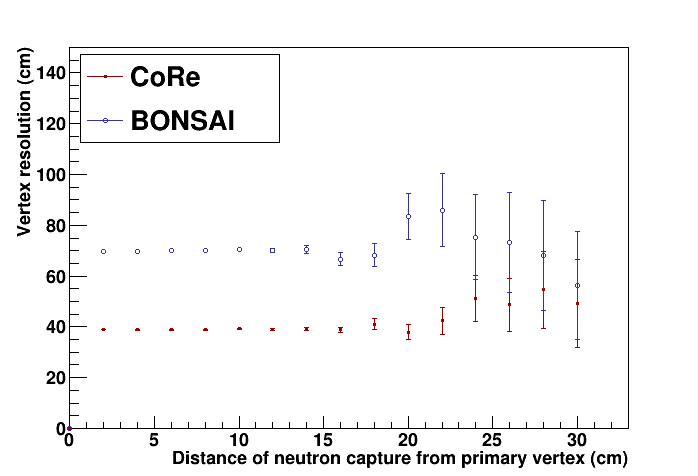}%
    }
    \caption{Comparison of vertex resolution for IBD positrons and neutrons in Gd-H$_2$O. Results for the 16~m [top and bottom left] and 22~m [top and bottom right] detectors, using the standard BONSAI reconstruction (circles) and CoRe (solid dots). Positron vertex resolution is plotted as a function of positron kinetic energy, while neutron vertex resolution is plotted as a function of distance of the neutron capture from the primary vertex. Vertex resolution is the distance from the true vertex within which 68\% of the events reconstruct. Note that very small errors are obscured by the markers in places.}
    \label{fig:vtxRes_water}
\end{figure}

\begin{figure}[htb]
    \centering
    \captionsetup[subfigure]{ width=0.5\textwidth}
    \subfloat[Positrons in the Gd-WbLS 16~m detector]{
    \includegraphics[width=0.5\textwidth]{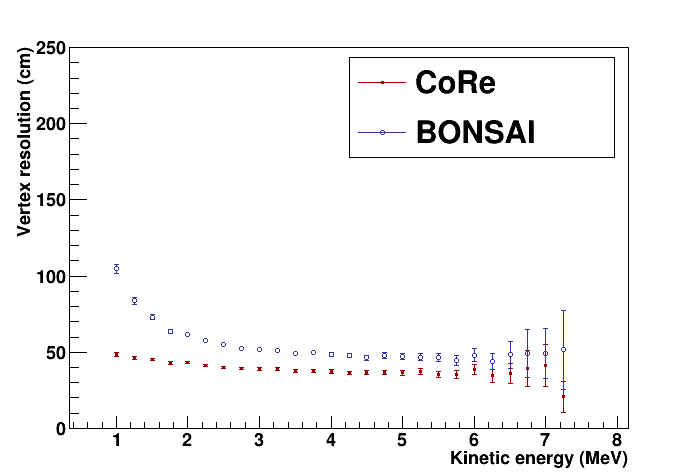}
    }
    \subfloat[Positrons in the Gd-WbLS 22~m detector]{
    \includegraphics[width=0.5\textwidth]{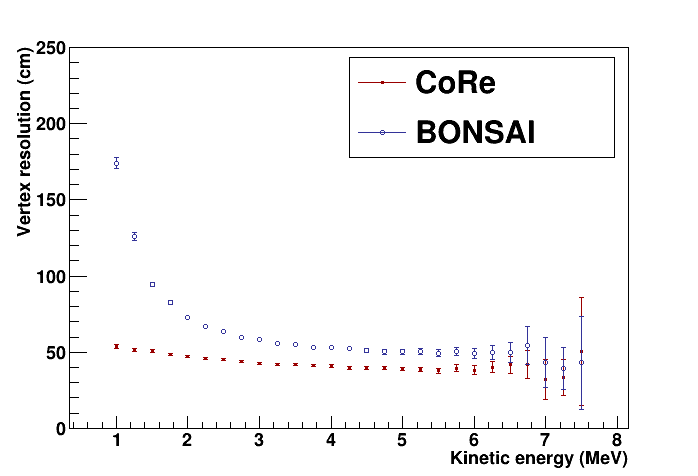}
    }
     \captionsetup[subfigure]{ width=0.5\textwidth}
    \subfloat[Neutrons in the Gd-WbLS 16~m detector]{
    \includegraphics[width=0.5\textwidth]{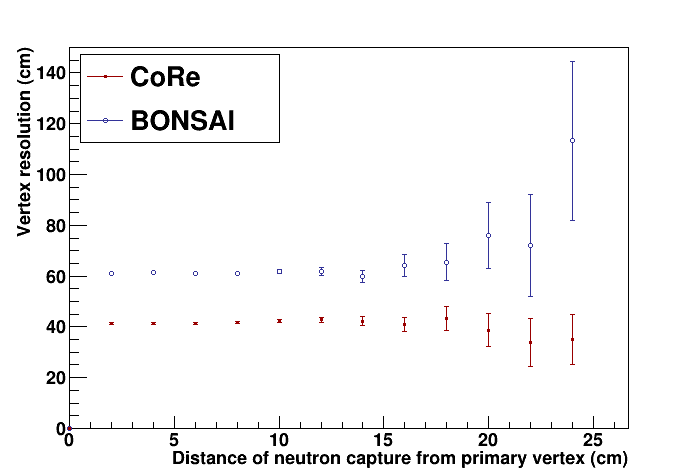}
    }
    \subfloat[Neutrons in the Gd-WbLS 22~m detector]{
    \includegraphics[width=0.5\textwidth]{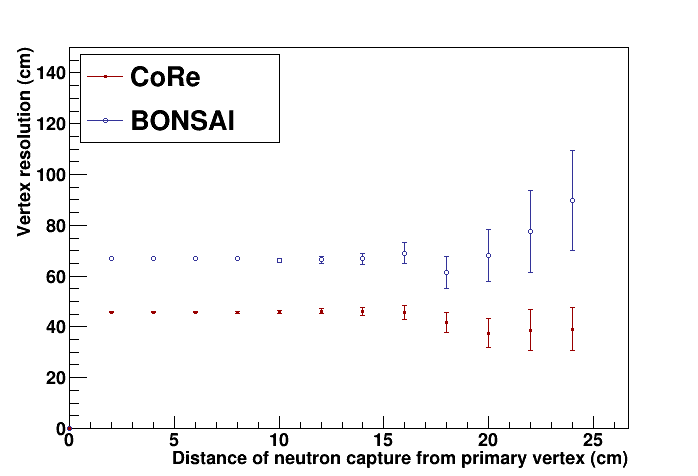}
    }
    \caption{Comparison of vertex resolution for IBD positrons and neutrons in Gd-WbLS. Results for the 16m [top and bottom left] and 22m [top and bottom right] detectors as a function of kinetic energy, using the standard BONSAI reconstruction (circles) and CoRe (solid dots). Vertex resolution is the distance from the true vertex within which 68\% of the events reconstruct. Note that very small errors are obscured by the markers in places.}
    \label{fig:vtxRes_wbls}
\end{figure}

CoRe improved the IBD vertex resolution compared to the BONSAI single-event fit for all detector configurations studied. Figures \ref{fig:vtxRes_water} and \ref{fig:vtxRes_wbls} show the results for Gd-H$_2$O and Gd-WbLS respectively for both the 16~m and 22~m detectors. The vertex resolution is expressed in terms of the distance from the true vertex within which 68\% of the events reconstruct. 

The improvement achieved using the coincidence reconstruction is particularly beneficial for positrons at lower energies in Gd-H$_2$O, where the low light yield from such events makes reconstruction without the additional light from the neutrons difficult. 

The vertex resolution for positrons is shown in Figures \ref{fig:vtxRes_water} and \ref{fig:vtxRes_wbls} as a function of positron energy and the results for 2.5~MeV and 5~MeV IBD positrons using BONSAI and CoRe are summarized in Table \ref{tab:vtxRes_summary}. The vertex resolution output by BONSAI improves with the addition of WbLS thanks to the additional, scintillation light but worsens with both BONSAI and CoRe with increasing detector size. Close to the peak of the positron signal, at 2.5~Mev, the resolution is improved by a factor of more than 2 in Gd-H$_2$O - from 84.0~cm to 41.3~cm in the 16~m detector and from 99.2~cm to 40.3~cm in the 22~m detector. In Gd-WbLS the vertex resolution at the same energy improved by more than 25\% from 54.8~cm to 39.9~cm in the 16~m detector and from 63.9~cm to 45.3~cm in the 22~m Gd-WbLS detector. The two methods tend to converge at the higher end of the energy range in the reactor IBD positron spectrum for all detector configurations. 

The vertex resolution for IBD neutrons is shown in Figures \ref{fig:vtxRes_water} and \ref{fig:vtxRes_wbls} as a function of the distance of the neutron capture from the primary vertex. It is expected that the resolution would increase (deteriorate) as the distance from the primary vertex increases. However, since the distances are within the vertex resolution achieved, and statistical errors are large, this effect is not significant. The vertex resolution for neutrons improved with CoRe by greater than 30\% at most distances in all configurations. Since we are reconstructing an effective neutron vertex from using the Cherenkov light resulting from the gammas emitted by the neutron capture, the case of perfect reconstruction (or zero distance between reconstructed and true neutron capture vertex) can only happen accidentally with this method.

\begin{table}[htb]
    \centering
        \captionsetup{width=\textwidth,font=small}
    \caption[Comparison of vertex resolution for IBD positrons]{Vertex resolution in cm with statistical error at selected energies for IBD positrons in the 16~m detector (top) and 22~m detector (bottom). A threshold timing goodness of 0.1 has been applied.}
    \begin{tabular}{c|c|c|c}
 Detector     &  Reconstruction  &  2.5~MeV                       & 5~MeV  \\
         \hline
16~m Gd-H$_2$O   & BONSAI      &  84.0 $\scriptstyle\pm0.8$    & 52.0 $\scriptstyle\pm1.0$   \\
              &CoRe         &  41.3 $\scriptstyle\pm0.3$    & 34.3 $\scriptstyle\pm0.4$  \\
        \rule{0pt}{4ex}
16~m Gd-WbLS  & BONSAI      &   54.8 $\scriptstyle\pm 1.1 $ &  47.2 $\scriptstyle\pm1.9$ \\
              & CoRe        &   39.9 $\scriptstyle\pm 0.5$  &  36.5 $\scriptstyle\pm0.9$ \\
         \rule{0pt}{4ex}     
22~m Gd-H$_2$O & BONSAI     & 99.2 $\scriptstyle\pm 1.0$    & 53.4 $\scriptstyle\pm1.1$ \\
               & CoRe       & 40.3 $\scriptstyle\pm0.2$     & 33.4 $\scriptstyle\pm0.4$  \\
        \rule{0pt}{4ex}       
22~m Gd-WbLS   & BONSAI     &  63.9 $\scriptstyle\pm 1.0$   &  50.4 $\scriptstyle\pm 1.7$ \\
               & CoRe       &  45.3 $\scriptstyle\pm 0.5$   &  39.0 $\scriptstyle\pm 1.0$ \\

    \end{tabular}
    \label{tab:vtxRes_summary}
\end{table}

\begin{figure}[htb]
    \centering
    \includegraphics[width=0.75\textwidth]{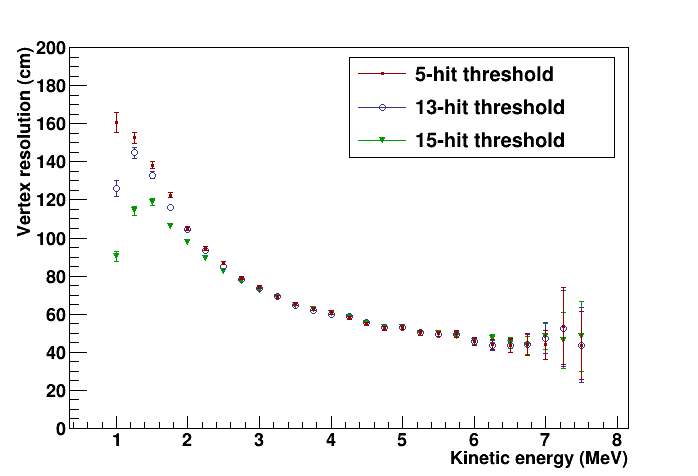}
    \caption{Effect of the hit threshold on the vertex resolution in the 16~m Gd-H$_2$O detector. Vertex resolution (distance from the true vertex within which 68\% of the events reconstruct) shown for inner-PMT hit thresholds of 5 hits (solid dots), 13 hits (circles) and 15 hits (triangles). As the threshold increases, the vertex resolution below 2.5~MeV - and in particular below 1.5~MeV - improves (decreases).}
    \label{fig:bonsai_threshold}
\end{figure}

The saturation of the positron vertex resolution below 1.5~MeV, which is observed in the single-event BONSAI reconstruction in the 16m Gd-H$_2$O detector, is due to the reconstruction threshold of 5 inner-PMT hits. This has the effect of improving the resolution at low energies as the threshold is increased (Figure~\ref{fig:bonsai_threshold}), since the hit threshold removes the events with the lowest light yields and is therefore more likely to remove events at the lowest energies. Events with a lower light yield are less likely to reconstruct well.

\begin{figure}[htb]
    \centering
    \captionsetup[subfigure]{width=0.5\textwidth}
    \subfloat[$\cos{\theta}$ as a function of true energy]{
    \includegraphics[width=0.5\textwidth]{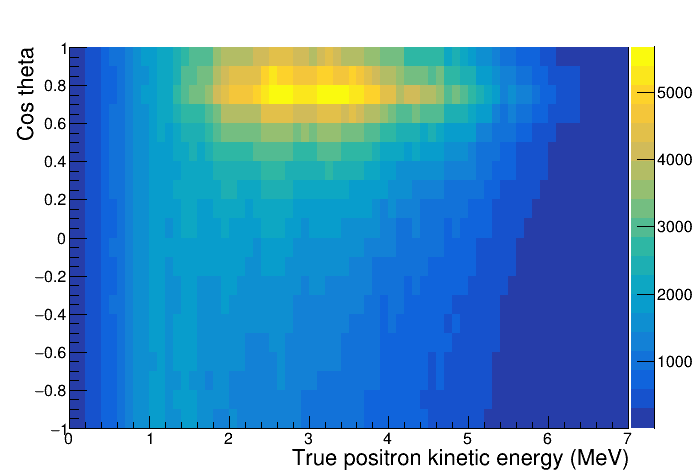}
    }
    \subfloat[Mean $\cos{\theta}$ as a function of true energy]{
    \includegraphics[width=0.5\textwidth]{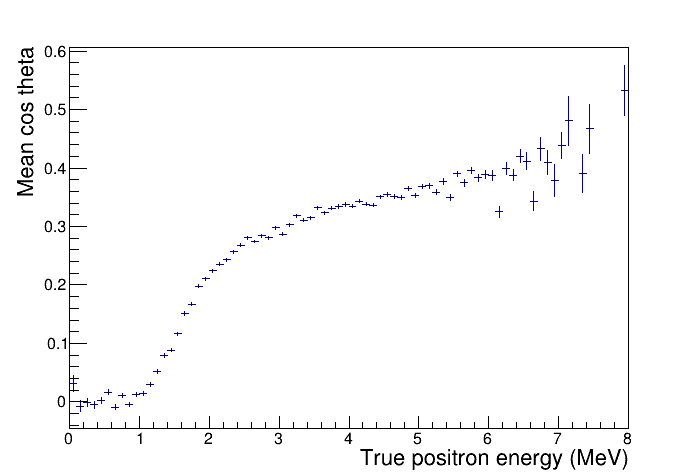}
    }
    \captionsetup{}
    \caption{Hit direction ($\cos{\theta}$) with respect to the positron direction as a function of true energy. The light from lower-energy particles is more isotropic (left) and the mean $\cos{\theta}$ between the particle direction and hit directions drops off steeply below $\sim$2.5~MeV, to around zero at 1~MeV (right). The BONSAI fit benefits from a higher fraction of backward hits.}
    \label{fig:cos_theta}
\end{figure}

The effect may be accentuated by the combination of the hit threshold with the greater isotropy of the light from the lowest-energy events (Figure~\ref{fig:cos_theta}). This helps to constrain the fit in the Cherenkov direction and to mitigate, to a degree, the difficulties presented by a relatively low light yield. 

Overall, these effects are diluted with CoRe and in Gd-WbLS due to the additional, isotropic light from the neutron capture and scintillation respectively.

With the addition of WbLS, the BONSAI results are significantly improved since more of the lower-energy positrons have sufficient light to achieve a reasonable fit. Although the benefit of CoRe over BONSAI is less marked in Gd-WbLS for this reason, the improvement in vertex resolution achieved with the coincidence reconstruction remains significant. It is slightly more significant in the larger tank at 2.5~Mev, which is where the BONSAI fit is least robust, despite the addition of the scintillation light. 

The vertex reconstruction worsens with increasing tank size with BONSAI for both detector fills. This is due in particular to the difficulty reconstructing vertices as the distance from the PMTs increases, making reconstruction towards the center of the detector more difficult as the tank size increases. This effect is much more marked with the BONSAI reconstruction in Gd-H$_2$O. 

The CoRe reconstruction offers the most improvement over BONSAI in both tank sizes with the Gd-H$_2$O medium. The CoRe results for the two detectors with this fill are consistent within the statistical error and may represent the limit of the reconstruction for detectors of this scale. The fit improves with the addition of WbLS in the 16~m detector and worsens with increasing tank size in Gd-WbLS, as expected. An unexpected result is the deterioration of the fit with the addition of WbLS in the 22~m detector. In fact, the addition of WbLS does not, overall, improve the vertex resolution with CoRe, compared to the vertex resolution achieved with CoRe in Gd-H$_2$O. Although there is improvement over the single-event fitting with CoRe, no gain in terms of the vertex resolution with CoRe is achieved by adding WbLS, which suggests there may be room for improvement.

\begin{figure}[htb]
    \centering
    \captionsetup[subfigure]{width=0.5\textwidth}
    \subfloat[Positrons, Gd-H$_2$O, 16m detector]{
    \includegraphics[width=0.5\textwidth]{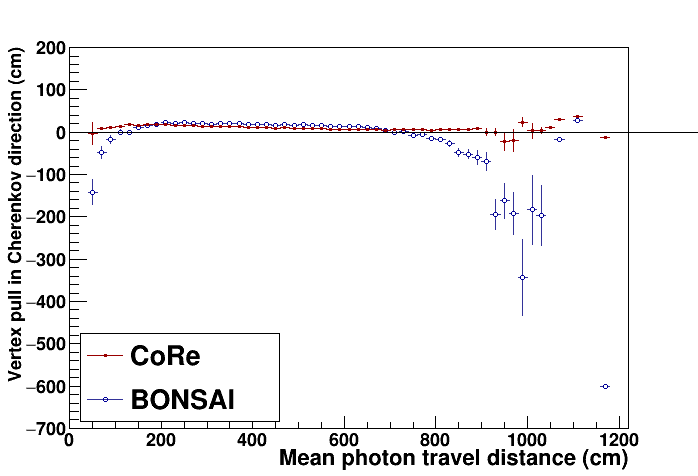}
    }
    \subfloat[Positrons, Gd-H$_2$O, 22m detector]{
    \includegraphics[width=0.5\textwidth]{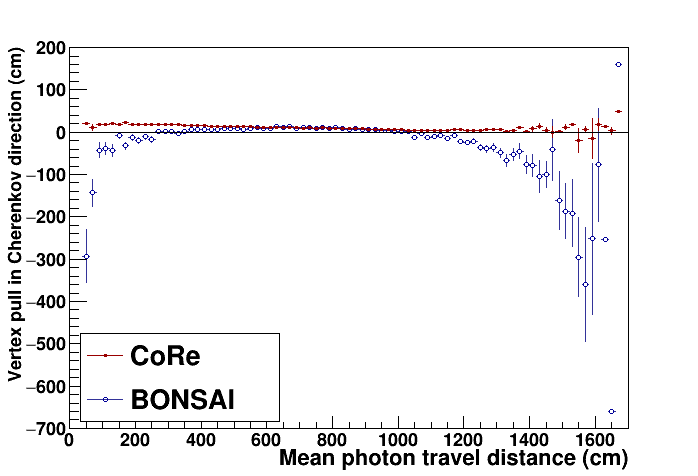}
    }
    \captionsetup[subfigure]{width=0.5\textwidth}

    \subfloat[Positrons, Gd-WbLS, 16m detector]{
    \includegraphics[width=0.5\textwidth]{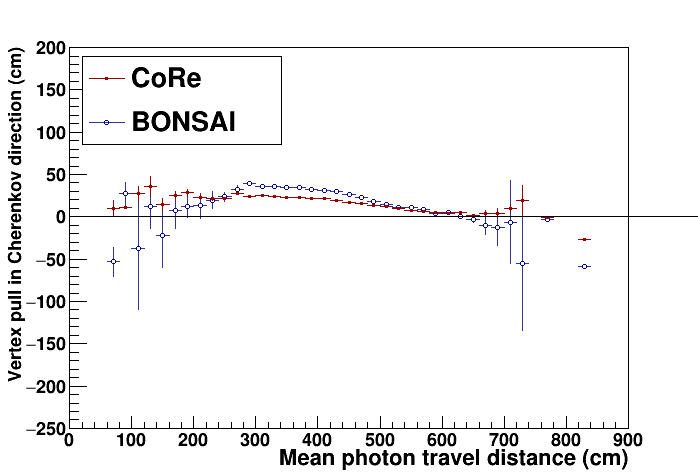}
    }
    \subfloat[Positrons, Gd-WbLS, 22m detector]{
    \includegraphics[width=0.5\textwidth]{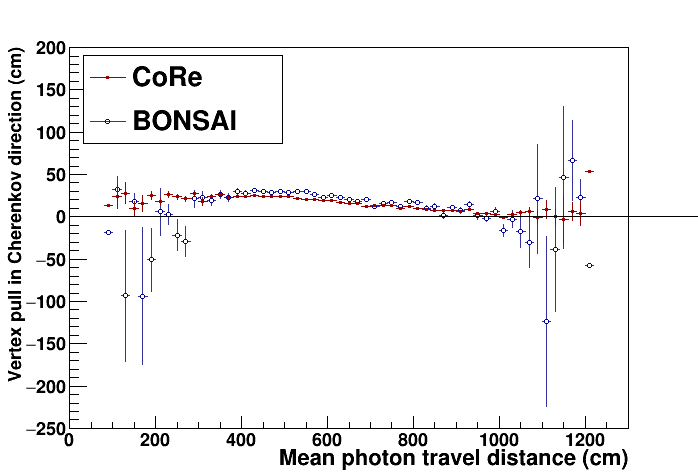}
    }
    \caption{Comparison of the pull in the Cherenkov direction for IBD positrons. Results for Gd-H$_2$O in the 16~m [top left] and 22~m [top right] detectors and in Gd-WbLS in the 16~m [bottom left] and 22~m [bottom right] detectors as a function of mean photon travel distance, using the standard BONSAI reconstruction (circles) and CoRe (solid dots). Note the different axis limits. Very small errors may be obscured by the markers in places.}
    \label{fig:pull}
\end{figure}

The fit is most difficult to constrain in the Cherenkov direction. This results in the \textit{pull} of the reconstructed vertex forwards or backwards along the Cherenkov direction with respect to the true vertex, which is shown in Figure \ref{fig:pull} as a function of mean photon travel distance. The mean photon travel distance is calculated for all of the hits in an event, excluding dark noise hits.

The \textit{pull} is particularly noticeable with the BONSAI single-event reconstruction at the extremes - these are events with vertices very close to the PMTs and very far from the PMTs. Although there is a forward pull over most of the distances, the most significant pull with the BONSAI reconstruction tends to be backwards along the Cherenkov direction and is worse in Gd-H$_2$O and in the larger detector.  

The additional light from the neutron helps the reconstruction to converge on a point in the Cherenkov direction and flattens the pull in the CoRe reconstruction in all detector configurations. The pull is close to zero (within $\sim$25~cm) at most mean photon travel distances in all detectors. Although the dispersion at the extremities is minimized with the CoRe reconstruction, there remains a consistent pull in the forward direction, which is also seen in the case of the BONSAI reconstruction except at the extremities. This suggests that there are potential improvements to be made. Incorporating charge information into the fit, for example, could offer improvements by accounting for multiply-hit PMTs.

\section{Background-Rejection Power of CoRe}
\label{core:tagging}

The coincidence reconstruction brings the additional benefit of powerful background rejection capability by providing a means by which to differentiate true, correlated pairs of events from false pairs - accidental coincidences of uncorrelated events. While reconstruction of true pairs results in a better vertex reconstruction, false pairs tend to result in a poorer reconstruction. Consequently, the timing goodness detailed in Section~\ref{core:goodness} can help to reject false pairs which otherwise pass coincidence cuts. 

\begin{figure}[htb]
    \centering
    \captionsetup[subfigure]{width=0.5\textwidth}
    \subfloat[Gd-H$_2$O, 16m detector]{
    \includegraphics[width=0.5\textwidth]{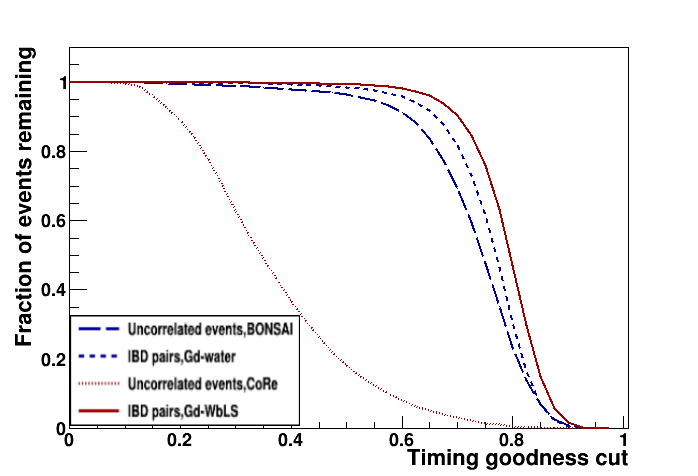}
    }
    \subfloat[Gd-H$_2$O, 22m detector]{
    \includegraphics[width=0.5\textwidth]{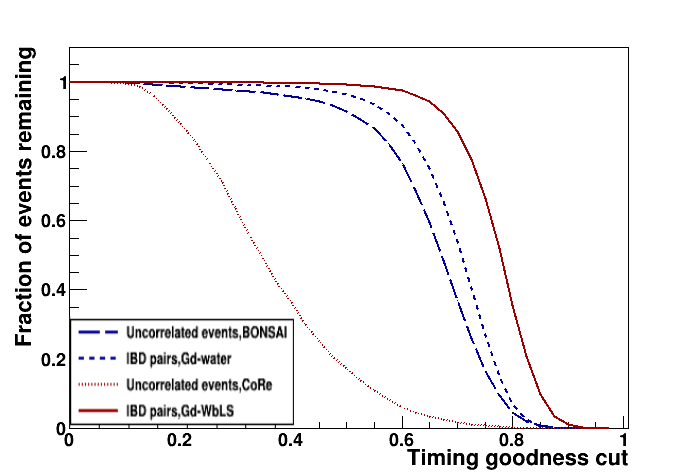}
    }
    \captionsetup[subfigure]{width=0.5\textwidth}
    \subfloat[Gd-WbLS, 16m detector]{
    \includegraphics[width=0.5\textwidth]{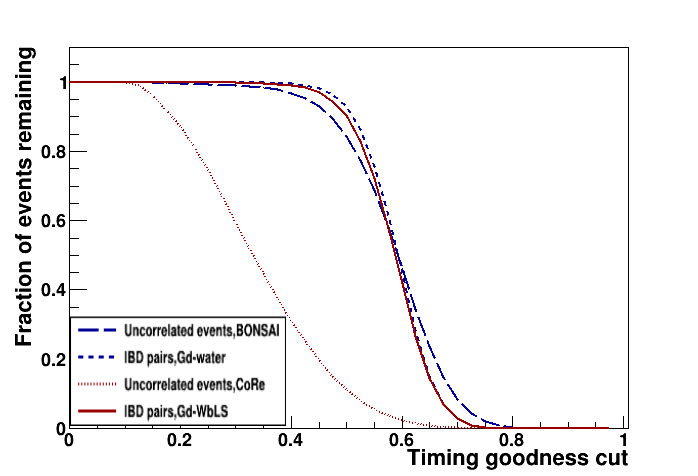}
    }
    \subfloat[Gd-WbLS, 22m detector]{
    \includegraphics[width=0.5\textwidth]{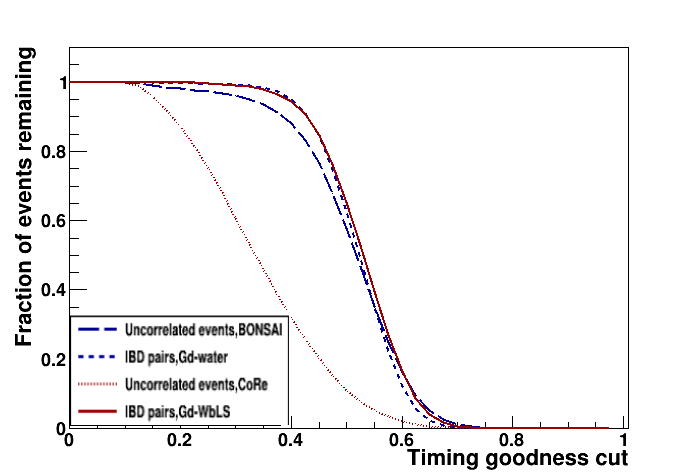}
    }
    \caption{Effectiveness of fit quality threshold for discriminating correlated and uncorrelated events. Fraction of correlated IBD pairs remaining as a function of a fit quality threshold, as measured by the timing goodness, applied to the prompt and delayed events in a pair for CoRe (solid) and BONSAI (short-dashed). The fraction of uncorrelated accidental coincidences remaining is shown for CoRe (dotted), with no other cuts applied, and for BONSAI (long-dashed), with an additional requirement that the events occur within 200$\mu$s of each other.}
    \label{fig:eff_goodness}
\end{figure}

Figure \ref{fig:eff_goodness} shows the fraction of true IBD pairs and accidental coincidences remaining as a function of a timing goodness threshold applied to both the prompt and delayed event in a pair. The results for BONSAI are shown on the plots for comparison. For the BONSAI results, a requirement that the time between events is less than $\rm 200\mu s$ has been applied to the uncorrelated pairs for consistency, but there is no cut on the distance between the prompt and delayed reconstructed vertices, which would normally be applied for background rejection with BONSAI in Gd-doped media\footnote{It is difficult to make a fair comparison of the overall background rejection capabilities of BONSAI and CoRe here, since there are a number of variables which can be used to reject background events. Cuts on these must be varied simultaneously and are dependent on the characteristics of both background and signal events.}.

The plots demonstrate the deterioration of the fit quality for uncorrelated events with CoRe, which is not reflected in the BONSAI results. Although the fit quality for uncorrelated events is better with BONSAI (\textit{i.e.}, the fit is not distorted by forcing a combined reconstruction), a threshold cut on the fit quality does not offer the potential for background rejection.

In all configurations, a threshold as low as 0.2 to 0.3 applied to the prompt and delayed event has the power to reject accidental coincidences of uncorrelated $\beta$ decays from ambient radioactivity in the detector and surroundings, which are a significant source of background in the reactor antineutrino range. Table \ref{tab:goodness_cuts} shows the fraction of events remaining as a function of timing goodness thresholds between 0.4 and 0.6 for both detector sizes and both media.

\begin{table}[htb]
    \centering
        \captionsetup{font=small}
    \caption{Fraction of IBD pairs and accidental coincidences remaining as a function of timing goodness (g) threshold in the four detector configurations.}
    \begin{tabular}{c|c c|c c|c c}
         
    Detector &  \multicolumn{2}{c|}{g $>$ 0.4} & \multicolumn{2}{c|}{g $>$ 0.5} & \multicolumn{2}{c}{g $>$ 0.6}  \\
    \hline
             &   IBD & acc                & IBD & acc                 & IBD & acc \\
         
         16~m Gd-H$_2$O  & 0.997 & 0.366  & 0.994  & 0.183  & 0.981 & 0.081  \\
         16~m Gd-WbLS    & 0.991 & 0.308 & 0.902 & 0.112 & 0.419  & 0.023  \\
         22~m Gd-H$_2$O   & 0.997 & 0.362 & 0.993 & 0.173  & 0.975  & 0.061  \\
         22~m Gd-WbLS    & 0.943  &  0.319  & 0.653  & 0.110  & 0.164  & 0.020  \\

    \end{tabular}
    \label{tab:goodness_cuts}
\end{table}

The cut is most effective in Gd-H$_2$O and a threshold of 0.6 would remove over 90\% of this type of accidental coincidence, while retaining $\sim$98\% of the signal events in the 16~m and 22~m detector. The timing goodness is not optimized for a Gd-WbLS fill, however there is still significant capacity for background rejection. A lower threshold of 0.4 would remove almost 70\% of the accidental coincidences while still retaining 99\% of the IBD pairs in the 16~m detector and 94\% of the IBD pairs in the 22~m detector.

Conversely, the timing goodness from the coincidence reconstruction can be helpful in selecting true pairs from data which is a combination of all types of events. This is important when high sample purity is important, or where coincident events form a background to a single-event signal.

\section{Conclusions and Future Work}
\label{sec:conclusions}

A new position reconstruction was implemented to take the light from two coincident events in a Cherenkov detector and reconstruct a combined vertex. This has been applied, using Monte Carlo simulations, to IBD pairs in the reactor antineutrino range in four detector configurations. The four configurations include two detector sizes - 16~m and 22~m - and two Gd-doped Cherenkov detection media - Gd-H$_2$O and Gd-WbLS. 

The new reconstruction improved the vertex resolution by a factor of more than 2 in the Gd-H$_2$O detectors for 2.5~MeV positrons, close to the peak of the reactor signal. The reconstruction improved by more than 25\% in the 16~m and 22~m Gd-WbLS configurations at the same energy and the vertex resolution for IBD neutrons improved by more than 30\% at most distances between the neutron capture and primary interaction vertex.

With position-based energy reconstruction, improved vertex resolution brings an improvement in energy resolution and the potential to see down to lower energies. A reconstruction threshold of 5 inner-PMT hits has been used in the BONSAI and CoRe implementations for this paper. The deterioration in the reconstruction at lower energies with the single-event reconstruction places a limit on the minimum number of hits required and how well the positions and energies can be reconstructed at the lowest energies. Given the stability of the fit with the coincidence reconstruction right down to the Cherenkov threshold, the reconstruction threshold would no longer be the limiting factor in seeing the lowest energy reactor antineutrinos and their energies could be reliably reconstructed.

A measure of the fit quality output by the coincidence reconstruction has been shown to be effective as a way to identify true pairs of correlated IBD events and reject false pairs. At the low energies of reactor antineutrinos, $\beta$-decay backgrounds from ambient radioactivity in and around a detector will contribute significantly to the rate of accidental coincidences of uncorrelated events, which can mimic the IBD signal in a Gd-doped medium. The power of CoRe to reject false pairs has been shown to help to reject 70\% to 90\% of this source of accidental coincidences while retaining $\sim$98\% or more IBD pairs. Other accidental coincidences \textit{e.g.}, of a signal event with a background event or of different types of background could be rejected in a similar way.

The coincidence reconstruction was developed to improve detection of antineutrinos from nuclear reactors for remote non-proliferation monitoring~\cite{kneale2021}. Monitoring reactor operation is essential to the verification of non-proliferation treaties and the reconstruction of IBD events is an important step towards making the monitoring of reactors via antineutrino detection possible. Remote monitoring with neutrinos is less intrusive than on-site monitoring and therefore may be more politically acceptable. This is discussed in more detail in~\cite{kneale2021}.
  
Implementation of the coincidence reconstruction for IBD in a gadolinium-doped detector could have wider application beyond reactor antineutrinos as the emerging Gd-H$_2$O and Gd-WbLS technologies are adopted. Super-K has already deployed gadolinium in its detector to three tenths of the planned final 0.1\% concentration. In Super-K, searches such as the hunt for supernova relic neutrinos and the detection of pre-supernova antineutrinos rely on IBD and the addition of gadolinium is seen as vital in these searches. Improving the vertex reconstruction and, perhaps more significantly, background rejection with an implementation of the coincidence reconstruction offers potential benefits in this area.

\section*{Acknowledgments}

The authors are grateful to M. Bergevin, for substantial contribution to the version of the RAT-PAC simulation used to generate the MC in this paper, to M. Askins, who wrote the generator used in the simulation of the coincident IBD events, and to Z. Bagdasarian, who incorporated the Gd-WbLS MC model into the RAT-PAC simulation.

Funding: This work was supported by the Atomic Weapons Establishment (AWE), as contracted by the Ministry of Defence, and the Science and Technology Facilities Council (STFC) in the UK, and the US Department of Energy's National Nuclear Security Administration.

\bibliographystyle{bibliography} 
\bibliography{bibliography}





\end{document}